\documentclass[sigconf]{acmart}
\usepackage{subcaption}
\usepackage[labelfont=up, textfont=up, font=small]{caption}
\usepackage{float}
\usepackage{graphicx}
\usepackage{enumitem}
\usepackage{textcomp}
\usepackage{url}
\usepackage{amsmath}
\usepackage{amsfonts}
\usepackage{multirow,rotating}
\usepackage{wrapfig}
\usepackage{tabularx}

\AtBeginDocument{%
  \providecommand\BibTeX{{%
    \normalfont B\kern-0.5em{\scshape i\kern-0.25em b}\kern-0.8em\TeX}}}

\usepackage{xspace}

\newcommand{\etc}{\emph{etc.}\xspace}
\newcommand{\ie}{\emph{i.e.}\xspace}

\newcommand{\etal}{\emph{et al.}\xspace}

\DeclareRobustCommand\textcmt[1]{{\fontfamily{cmtt}\selectfont{#1}}}

\newcommand{\bi}{\textbf{\textit{BI}}}

\usepackage{fixltx2e}
\def\SPSB#1#2{\rlap{\textsuperscript{{#1}}}\SB{#2}}

\def\SB#1{\textsubscript{{#1}}}

\definecolor{our_red}{HTML}{f84030}
\definecolor{our_blue}{HTML}{1f77b4}
\definecolor{our_green}{HTML}{037815}

\begin{document}

\fancyhead{}

\title{A Peek into the Political Biases in Email Spam Filtering Algorithms During US Election 2020}

\renewcommand{\shortauthors}{et al.}

\author{Hassan Iqbal, Usman Mahmood Khan, Hassan Ali Khan, Muhammad Shahzad}
\affiliation{%
	\institution{Department of Computer Science, North Carolina State University}
	\city{Raleigh}
	\state{North Carolina}
	\country{USA}
}
\email{{hiqbal,ukhan3,hakhan,mshahza}@ncsu.edu}

\begin{abstract}
	Email services use spam filtering algorithms (SFAs) to filter emails that are unwanted by the user.
	However, at times, the emails perceived by an SFA as unwanted may be important to the user.
	Such incorrect decisions can have significant implications if SFAs treat emails of user interest as spam on a large scale.
	This is particularly important during national elections.
	To study whether the SFAs of popular email services have any biases in treating the campaign emails, we conducted a large-scale study of the campaign emails of the US elections 2020 by subscribing to a large number of Presidential, Senate, and House candidates using over a hundred email accounts on Gmail, Outlook, and Yahoo.
	We analyzed the biases in the SFAs towards the left and the right candidates and further studied the impact of the interactions (such as reading or marking emails as spam) of email recipients on these biases.
	We observed that the SFAs of different email services indeed exhibit biases towards different political affiliations.
	We present this and several other important observations in this paper.
\end{abstract}


\keywords{US Elections, 
Emails, 
Spam, 
Bias,
Political Bias,
Algorithm Bias}

\settopmatter{printacmref=false}
\renewcommand\footnotetextcopyrightpermission[1]{}

\maketitle

\section{Introduction} \label{lab:introduction}
The spam filtering algorithms (SFAs) in the widely-used email services of today such as Gmail, Outlook, and Yahoo do not provide any transparency on their internal workings. 
Given the lack of this transparency, an important question to study is whether these SFAs hold any biases towards certain political affiliations. 
This question is motivated by the growing body of evidence suggesting that the biases in online algorithms can influence undecided voters.
For example, Epstein $\etal$ showed that the bias in search rankings can shift the voting preferences of the undecided voters by as much as 20\% without those voters being aware of the manipulation \cite{undecided_voters}.
Furthermore, several US political candidates in the 2020 US election raised concerns that the email clients were filtering out the campaign emails they were sending to their constituents \cite{gopsundar}.

\noindent \textbf{Research Questions:} 
In this paper,
\let\thefootnote\relax\footnote{\textcolor{red}{This is a pre-print. The final version is published in The ACM Web Conference 2022 (WWW '22). For the published version and citation, please refer to DOI: https://doi.org/10.1145/3485447.3512121}}we attempt to assess the fairness of the SFAs of three dominant email services, Gmail, Outlook, and Yahoo, in the context of the 2020 US election.
Specifically, we study the following four research questions:

\begin{itemize}
	\item \textbf{Q1:} Do SFAs of email services exhibit aggregate political biases?
	How do these biases compare across email services?
	\item \textbf{Q2:} Do SFAs treat similar emails from senders with different political affiliations in the same way?
	\item \textbf{Q3:} Do the interactions of the users with their email accounts, such as reading emails, impact the political biases of SFAs? 
	\item \textbf{Q4:} Do SFAs exhibit different political biases for recipients belonging to different demographic?
\end{itemize}

\noindent To the best of our knowledge, there is no prior published work that has examined biases in SFAs towards political campaigns.
\\


\noindent \textbf{Proposed Methodology:} To answer these questions, we conducted an extensive study during the 2020 US election over a period of 5 months from July 1, 2020 to November 30, 2020 on Gmail, Outlook, and Yahoo. 
We created 102 email accounts and subscribed to 2 Presidential, 78 Senate, and 156 House candidates.
To accurately estimate the political biases and mitigate any potential effects of demographics (ethnicity, age, and gender), we created multiple email accounts with different combinations of demographic factors and designed two experiments. 
The first experiment studies the general trends of biases in SFAs across the email services for the Presidential, Senate and House candidates.
The second experiment studies the impact of different email interactions such as reading the emails, marking them as spam, or vice versa on the biases in SFAs. 
We designed an automated process to perform all the subscriptions, and took periodic backups to keep all the email accounts active as well as to keep track of the correct number of spam emails received over the course of data collection for each of the three services.

We made several important observations in our study. 
For example, as an \textit{aggregate trend}, Gmail leaned towards the left while Outlook and Yahoo leaned towards the right. 
Yahoo retained about half of all the political emails in inbox (up to 55.2\% marked as spam) while outlook filtered out the vast majority of emails (over 71.8\%) from all political candidates and marked them as spam.
Gmail, however, retained the majority of left-wing candidate emails in inbox ($<10.12\%$ marked as spam) while sent the majority of right-wing candidate emails to the spam folder (up to 77.2\% marked as spam).
We further observed that the percentage of emails marked by Gmail as spam from the right-wing candidates grew steadily as the election date approached while the percentage of emails marked as spam from the left-wing candidates remained about the same. 
We present these and several other important observations in this paper.

\noindent \textbf{Key Contributions:} 
\begin{itemize}
    \item To the best of our knowledge, this paper presents the first study that extensively explores the political biases in SFAs.
	\item We used the propensity score matching approach \cite{rosenbaum1983} to determine whether the SFA of any given email service 
	provided same treatment to similar emails from candidates of different political affiliations.    
	\item We have aggregated and analyzed a large data set of over 318K political emails across the three email services. This data set is available at \cite{datasetgithub}.
\end{itemize}

\vspace{-0.1in}
\noindent \textbf{Paper Organization:}
Next, we discuss the related work in $\S$ \ref{lab:related_work} and describe our methodology in $\S$ \ref{lab:methodology}.
We then present the extensive analysis of our data set in $\S$ \ref{lab:analysis_v3} to study our four research questions.
Finally, in $\S$ \ref{sec:ConcludingDiscussion}, we summarize our observations, discuss their implications and provide some suggestions, and conclude the paper.

\vspace{-0.1in}
\section{Related Work} \label{lab:related_work}
Researchers have proposed several different definitions of spam emails.
Butler defines spam as any unsolicited email that comes from an entity that the recipient is not already aware of or has no interest in knowing about \cite{butler_spam}.
Cormack \textit{et al} similarly defines spam as any unsolicited or unwanted email that is sent indiscriminately, and has no current relation to the recipient \cite{cormack_spam}.
Similar definitions have been proposed in other related articles \cite{androu_spam} \cite{sphamus_spam}. 
In contrast, Google defines spam as any content that is unwanted by the user \cite{google_spam}. 
This is significantly different from the criteria proposed by the previous research in that the spam email does not have to meet any of the explicitly defined conditions so long as there is a reason to believe that the email may be unwanted by the recipient \cite{google_spam}. 
Other email clients such as Outlook and Yahoo have not made their definitions of spam public. 
In this study, our objective is to study how these email services treat emails from the political websites that the recipient has subscribed to, and if that treatment has any biases, irrespective of how these services define spam. 

One prior work examines the manipulative tactics in political emails~\cite{elec2020}. 
The focus of this work is on finding different manipulative tactics that the campaigns use to encourage readers to open emails and donate. 
It categorizes manipulative tactics as  click baits (forward referencing, sensationalism, urgency), and user interface manipulation (obscured names, ongoing thread, subject manipulation).
To the best of our knowledge, there is no past published work on examining the political biases in the SFAs of different email services.
However, prior work has examined the biases of several other web-based algorithms. 
For example, the past studies have demonstrated that the online advertising results may be personalized \cite{online_1} \cite{online_2}. 
Hannak \textit{et al} found significant personalization in Google web search based on the account login status and the geolocation of the IP address \cite{hannak_16}. 
Another study reported that 15.8\% of the Bing search results were personalized for different users \cite{hannak_18}. 
Puschmann analyzed Google search results for German political parties and candidates and found that the results were personalized based on several factors including location, language, and time \cite{puschmann}. 
Huyen \textit{et al} showed that these results are further personalized based on the user's browsing history \cite{huyen}. 
However, to the best of our knowledge, no such study exists for email SFAs in the context of political campaign emails.

\section{Methodology} \label{lab:methodology}

\subsection{Emails and Demographics}
\label{subsec:EmailsAndDemographics}
We used three email services, Gmail, Outlook, and Yahoo, and created 102 email accounts, 34 on each of the three services.
To accurately estimate the political biases and mitigate the potential effects of demographic factors such as ethnicity and age, we created our email accounts with different combinations of these factors. 
As email services do not explicitly collect ethnicity information, we assigned a different name to each email account that we randomly picked from a database of common names associated with White, African-American, Hispanic, Asian, and South Asian ethnicities. 
For age, we assigned each email account to one of the three age groups of 18-40, 40-65, and 65+. 
Finally, we randomly assigned male and female genders to the email accounts. 
To conform to the ethical standards, none of the email accounts that we created belonged to any real users and all the accounts are new with no prior history.
We manually created all the accounts by following the account creation procedure of the three email services.

\subsection{Subscribed Candidates}
We subscribed our email accounts to presidential, Senate, and House candidates, described next. 

\subsubsection{Presidential Candidates:} 
This category includes the two Presidential candidates, one from the left, \ie, Joe Biden (Democrat), and one from the right, \ie, Donald Trump (Republican). 

\subsubsection{Senate and House Candidates:}
\begin{figure}[b]
	\vspace{-0.125in}
	\begin{center}
		{\includegraphics[width=0.7\columnwidth]{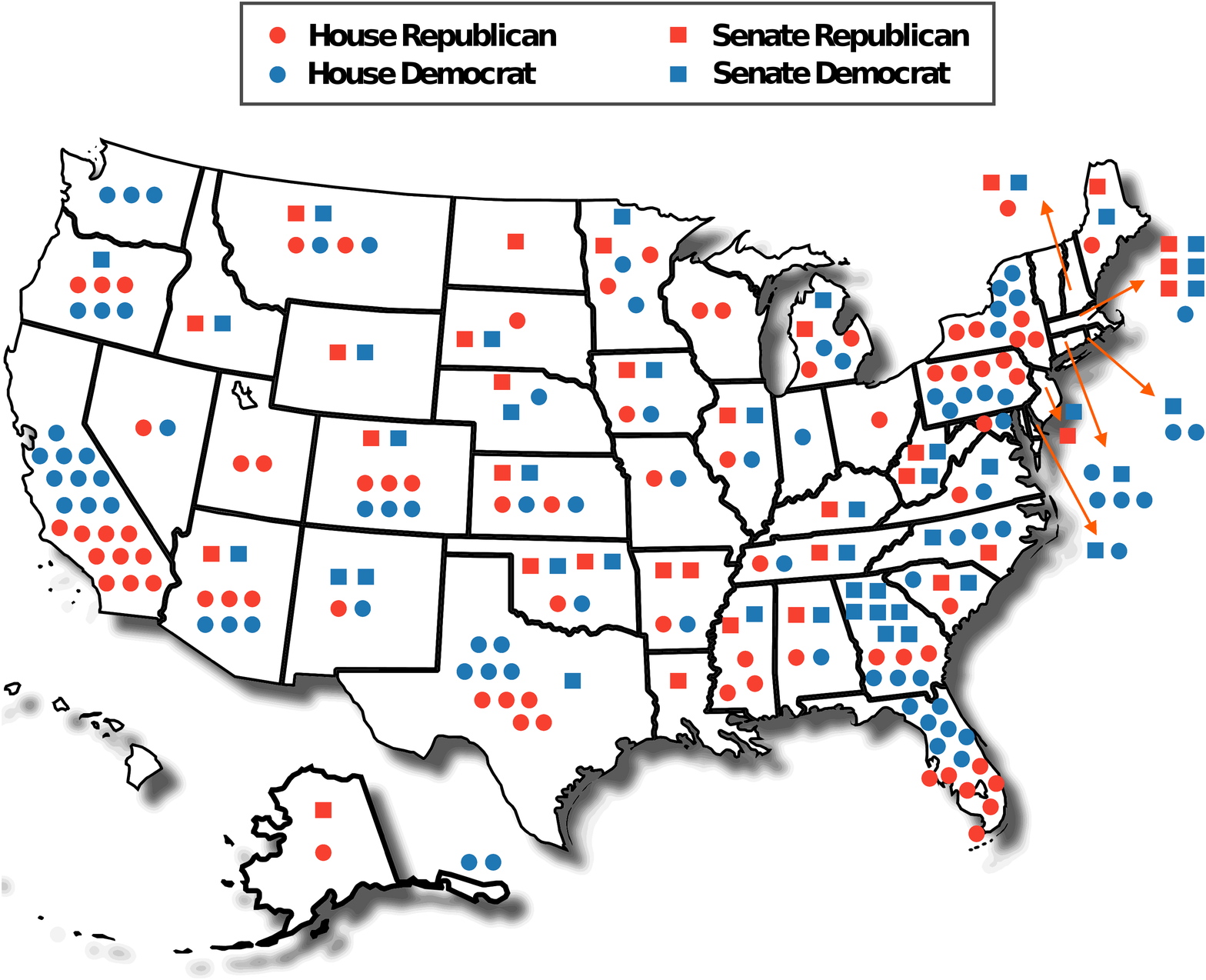}}
	\end{center}
	\vspace{-0.2in}
	\caption{Distribution of Senate \& House Candidates in our subscriptions.}
	\label{fig:map}
	\vspace{-0.075in}
\end{figure}

This category combines candidates from both the United States Senate and the House of Representatives, as shown in Fig. \ref{fig:map}.
The blue and red circles represent the left and right candidates, respectively, in the US House of Representatives that we picked for this study.
Similarly, the blue and red squares represent the left and right Senate candidates.
In some states, we subscribed to a different number of left and right candidates due to four reasons.
First, different states have different number of seats in the House depending on factors like the population of the state.
Second, each Senate and House election is contested by different number of candidates from both Democratic (left) and Republican (right) parties. 
Third, some candidates only had government affiliated websites (with \textit{.gov} domain names), which are prohibited to send out campaign emails \cite{officalCiteUse, senateEthics}.
Fourth, due to our automated subscription methodology (will be described shortly), we could not subscribe to the campaign websites that required filling out a CAPTCHA form.
We initially chose the campaign websites of all the left and the right Senate and House candidates in the 50 states.
After filtering those websites based on the third and the fourth reasons stated above, we ended up with an unequal number of campaign websites of the left and the right candidates, in various states, for both the Senate and the House.

To reduce the gap between the number of left and right candidates, our subscription methodology was as follows. 
If any state had more than one but unequal number of left and right candidates, we subscribed to the maximum number of candidates such that the counts of the left and right candidates were the same. 
To keep as many states in our analysis as possible, we did not use this approach in states like Alaska (with only 1 Republican senate candidate), where we found candidates with non-government affiliated websites from only one of the parties.
There were 11 such states.
As a result, there is a small difference between the counts of the left and the right House and Senate candidates in our subscriptions.
In total, we subscribed to Senate candidates from 36 states with 78 subscriptions (44 L and 34 R), and House candidates from 42 states with 156 subscriptions (81 L and 75 R).
With both Senate and House subscriptions combined, we were able to cover all 50 states.

\vspace{-0.05in}
\subsection{Experiment Design}
To answer our research questions, we designed two experiments, described next.

\subsubsection{Baseline Experiment (E1)}
This experiment brings forth the true trends, which have not been subjected to any personalizations, of the biases in the SFAs of the three email services.
The observations from this experiment also serve as the baseline for the comparison of observations from the next experiment.
This experiment involved 66 email accounts, 22 accounts per service.
For each email service, we assigned 6 accounts to White Americans, 6 to African Americans, 4 to Hispanic Americans, 3 to Asian Americans, and 3 to South Asian Americans.
Among each set of 22 email accounts, the three age groups of 18-40, 40-65, and 65+ had 8, 8, and 6 accounts, respectively.
Each email account subscribed to all of the Presidential, Senate, as well as House candidates included in our study.
We kept the email accounts in the baseline experiment untouched and did not subject them to any interactions.

\subsubsection{Interaction Experiment (E2)} \label{subsubsec:e2}
This experiment studies the impact of different interactions with the email accounts on the biases in SFAs. 
It contains 12 $\times$ 3 email accounts that subscribe to all the Presidential, Senate, and House candidates in our study.
In this experiment, we randomly assigned ethnicity, age, and gender to the 12 email accounts of each service.
We split this experiment into three groups containing 4 $\times$ 3 accounts each, and performed three different interactions, one per group.
These three interactions included reading all emails, moving all emails from inbox to spam folder, and moving all emails from spam folder to inbox.
We chose these three interactions for four reasons.
First, reading is the most common action that one performs on an email of interest.
Second, moving emails from inbox to spam folder reflects the user's preference that the user is no longer interested in such emails.
Third, moving emails from spam folder to inbox reflects the user's preference that the user is interested in such emails, and the SFA wrongly marked them as spam.
Fourth, we choose to move all emails instead of randomly selecting them to avoid introducing any bias based on the content or sender of the email.
We performed the reading interaction on $4\times3$ email accounts every 24 hours, and the other two interactions of moving from inbox to spam and spam to inbox every 5 days on their respective $4\times3$ email accounts.

\vspace{-0.05in}
\subsection{Subscription Process}
As each email account subscribes to 236 websites (2 Presidential, 78 Senate, and 156 House) on average, we had to complete 24,072 subscriptions.
To automate the subscription process, we wrote a Python script that scrapes websites using Selenium library \cite{selenium}, and automatically fills out the subscription form to subscribe email accounts to the campaign websites.
However, since there are 236 unique websites in our experiments, it was not feasible to write separate scraping codes for each of these websites.
To address this challenge, we developed a general algorithm that extracts all forms from a website, determines if one or more of these forms are related to subscription, and then fills them out.
To ensure that we did not miss any subscriptions, we generated the logs for the failed subscription attempts that we later completed manually by visiting the corresponding websites.

\vspace{-0.05in}
\subsection{Data Set}
We started our data collection on July 1\textsuperscript{st}, 2020 and ended it on February 28\textsuperscript{th}, 2021.
However, we observed that the volume of emails from the campaign websites significantly dropped after November 20\textsuperscript{th}, 2020.
Therefore, we truncated the data set on November 30\textsuperscript{th}, 2020 and conducted analysis on emails that we collected over these 5 months (153 days).
We collected 318,108 emails across the three services.
The content that we have collected for each email contains mail header fields such as 
\textcmt{Subject}, 
\textcmt{From}, 
\textcmt{To}, 
\textcmt{Date}, 
\textcmt{Message-ID}, 
\textcmt{Delivered-To}, 
\textcmt{Received-SPF}, 
\textcmt{Received-by},\\
\textcmt{Content-Type}, 
\textcmt{MIME-Version}, 
\textcmt{Content-Type}, 
and message body.

\vspace{-0.05in}
\section{Analysis} \label{lab:analysis_v3}
In this section, we study the biases in the SFAs of Gmail, Outlook, and Yahoo.
In the following four subsections, we present our results and observations to answer the four questions mentioned in $\S$ \ref{lab:introduction}.

\vspace{-0.05in}
\subsection{Political Biases in Spam Filtering Algos.} \label{subsec:polticalBiasE1}
In this section, we study whether the SFAs exhibit political biases and how these biases compare across different email services.
Specifically, we examine whether the SFAs in each of the three email services:
1) lean towards the right by sending more campaign emails of the left to the spam folder,
2) lean towards the left by sending more campaign emails of the right to spam, or 
3) remain neutral by giving similar treatment to the campaign emails from both left and right.
We answer this question by analyzing the campaign emails of the Presidential, Senate, and House candidates that we received in the 22 accounts of each email service in the baseline experiment (E1).
We first analyze the aggregate political bias in the SFAs in terms of the percentage of the left and the right campaign emails that are marked as spam.
After that, we conduct a temporal evaluation of the political bias at weekly intervals.
Last, we analyze the biases in the SFAs for the campaign emails from individual Senate and House candidates.

\vspace{-0.05in}
\subsubsection{Aggregate Political Bias} \label{subsubsec:aggregatePolBias}
\begin{figure}[htbp]
	\vspace{-0.15in}
	\begin{center}
		{\includegraphics[width=1\columnwidth]{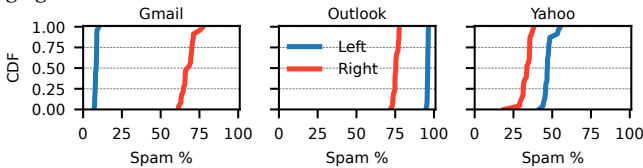}}
	\end{center}
	\vspace{-0.15in}
	\caption{Cumulative distribution of the \%age of left (blue) and right (red) emails marked as spam in each of the 22 email accounts of each service.}
	\label{fig:dist_e1}
	\vspace{-0.175in}
\end{figure}

We observed that the SFAs of the email services indeed exhibit political biases: they treat the left and the right campaign emails differently.
Gmail leans towards the left as it marks a higher percentage of the right emails as spam.
Outlook and Yahoo, on the other hand, lean towards the right.
Each blue line in Fig. \ref{fig:dist_e1} shows the cumulative distribution (CDF) of the percentage of left emails marked as spam in each of the 22 email accounts of the corresponding email service.
The red lines show the same for the right emails.
We observe that each CDF line rises rapidly, which demonstrates that the SFA of each email service is fairly consistent across the 22 email accounts in its treatment of emails as spam.
We further observe that Gmail marks a significantly higher percentage (67.6\%) of emails from the right as spam compared to the emails from left (just 8.2\%).
Outlook is unfriendly to all campaign emails, more unfriendly to the left than to the right.
It marks a higher percentage of left (95.8\%) emails as spam than those of right (75.4\%). 
Yahoo marks 14.2\% more left emails as spam than the right emails.
Each of these numbers above represents the average across the 22 accounts of the corresponding services.
Onward, we will refer to these observations about Gmail leaning towards the left and Outlook and Yahoo towards the right as the \textit{aggregate trend}.

\subsubsection{Temporal Evolution of Political Bias}
\begin{figure}[htbp]
	\vspace{-0.15in}
	\begin{center}
		{\includegraphics[width=1\columnwidth]{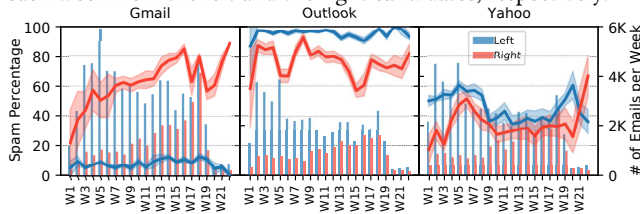}}
	\end{center}
	\vspace{-0.175in}
	\caption{Left y-axis: average (line plot) and standard deviation (shaded bands) of the \%age of emails marked as spam each week. Right y-axis: \# of left and right emails (vertical bars) received each week.}
	\label{fig:e1_weekly}
	\vspace{-0.15in}
\end{figure}

Next, we discuss
1) whether the spam percentage varies over time for the left and right campaign emails, and 
2) whether there is any correlation between the spam percentage and the number of received emails.
We present our observations using Fig. \ref{fig:e1_weekly}.
The blue and the red solid lines in this figure show the average percentage (averaged across the 22 email accounts) of the left and the right emails, respectively, marked as spam each week.
The shaded bands around the solid lines show the standard deviation for each week. 
The blue and the red vertical bars show the total number of emails (inbox + spam) received each week from the left and the right candidates, respectively.

\begin{figure}[htbp]
	\vspace{-0.15in}
	\begin{center}        
		{\includegraphics[width=1\columnwidth]{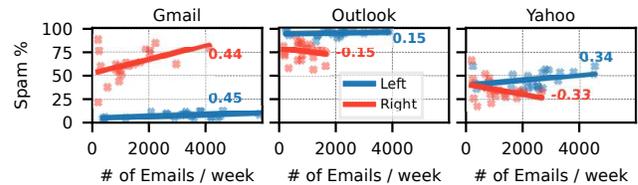}}
	\end{center}
	\vspace{-0.16in}
	\caption{Pearson correlation between the weekly average spam \%age and the weekly \#of emails in the 22 accounts.}
	\label{fig:e1_weekly_corr}
	\vspace{-0.14in}
\end{figure}

We make three important observations from Fig. \ref{fig:e1_weekly}.
First, in Gmail, we observe an increasing trend in the right spam percentage over time, whereas the left spam percentage did not vary much and remained under 15\%.
For the right spam, we also observe that the spam percentage increased with an increase in the number of right campaign emails.
The left spam emails also show this trend, albeit, it is less apparent in Fig. \ref{fig:e1_weekly}.
Thus, we plot Fig. \ref{fig:e1_weekly_corr}, where we show the Pearson correlation ($r$) of the number of emails from left and right received in a week to the percentage of the left and right emails marked as spam in that week, respectively, in the the 22 email accounts of each service.
We observe from this figure that Gmail has a positive correlation for both the left ($r=0.45$) and the right ($r=0.44$).
This shows that Gmail marks a larger fraction of emails as spam as the volume of emails increases. 
Second, Outlook is almost indifferent to the left spam percentage as it stayed above 85\% for all the weeks.
The variation in the number of left emails did not cause noticeable variation in the spam percentage ($r=0.15$).
The right spam percentage shows more volatility over the 22-week period, but still, there is no correlation ($r=-0.15$) between the right spam percentage and the number of emails from the right.
Third, in Yahoo, the spam emails for both the left and the right initially increased and then decreased over time for most of the weeks.
With the exception of the last two weeks, when the elections were over and the candidates significantly reduced the number of emails, the spam percentage for the left emails remained higher than that of the right emails.
For the right candidates, the spam percentage decreased with the increase in the number of right emails ($r=-0.33$).
Contrarily, the left spam percentage increased with the increase in the number of left emails ($r=0.34$).

To summarize, the \textit{aggregate trend} that we observed in \S \ref{subsubsec:aggregatePolBias} holds over weekly interval as well, $\ie$ Gmail leans towards the left whereas, Outlook and Yahoo towards the right.
In Gmail and Yahoo, the number of emails from the left and from the right have a noticeable influence on the percentages of their emails marked as spam.
However, such influence is not seen in the case of Outlook.

\vspace{-0.06in}
\subsection{Impact of Political Affiliation}\label{subsec:ImpactofPoliticalAffiliation}
\vspace{-0.01in}
In \S \ref{subsec:polticalBiasE1}, we saw that there are indeed aggregate biases in the SFAs of different email services.
However, an important question still remains: 
\textit{do these biases exist even when we consider only those emails from the left and the right candidates that have very similar attributes?}
In other words, does the political affiliation of the sender alone play a significant role in getting an email marked as spam?
An answer in affirmative would be worrisome because a sizable chunk of voting population heavily relies on these email services, and these biases could sway their decisions about who to vote for and whether to even cast their votes.
There is growing evidence that the online interactions of people shape their political opinions.
For example, Hu $\etal$ showed that custom Google search snippets consistently amplify political bias \cite{snippet}.

An ideal way to obtain the answer to this question would be to have both the left and the right candidates send the same set of emails to our email accounts.
Then, by comparing the percentage of those emails marked as spam when sent by the left candidates with the percentage of those same emails marked as spam when sent by the right candidates, we could answer this question.
While ideal, unfortunately, this approach is impractical because it was beyond our control to decide what emails different candidates sent.
However, due to the large volume of emails in our data set, it is still possible to obtain approximately the same effect as the ideal method described above using the well-known statistical method of propensity score matching (PSM) \cite{psmbook}.

PSM is a popular statistical method that is used to preprocess data from observational studies in which it is not feasible to conduct a randomized controlled trial, \ie, the studies where it is not possible to control who gets to be the member of the treatment group and who gets to be the member of the control group.
PSM essentially takes various attributes, commonly known as \textit{covariates}, of the members of the treatment and the control groups and selects appropriate members from the two groups to create a new treatment and a new control group, also known as \textit{matched} groups, such that the distribution of any given covariate of the members of the matched treatment group is similar to the distribution of that covariate of the members of the matched control group.
As a result of this similarity of distributions, the observations that one makes from the two matched groups about the effects of the treatment can be approximated to be the observations from a randomized controlled trial \cite{psmbook}.
Note that the original treatment and control groups are also known as \textit{unmatched} groups.
We have provided a quick primer on PSM in the supplementary material, \S \ref{subsubsec:PSM--A Primer}.
For more details on PSM, we refer the interested readers to \cite{rosenbaum1983,psmbook,psmguidance}.

Next, we first map our emails problem to PSM and describe the covariates that we have selected.
After that, we present our observations from the matched groups that PSM creates and study whether significant biases exist in SFAs even in the matched groups.

\subsubsection{Mapping Emails Problem to PSM}
\label{subsubsec:mappingEmailstoPSM}
PSM takes all the emails, selects a subset of emails from them based on the values of appropriate covariates (which we will discuss shortly), and creates the two \textit{matched} groups.
All the emails in one group are from the left and the other are from the right.
An important property of these groups is that each email in one group has a corresponding email in the other group such that the values of the covariates of the two emails are very similar.
Thus, the emails in the two matched groups are very similar in terms of the selected covariates.
Once PSM creates the matched groups, we can then study whether or not the SFA of any given email service marked a comparable percentage of the emails in the two matched groups as spam.

We applied PSM on our email data set collected during the baseline experiment (E1).
In our application of PSM on emails in any given email service, the unmatched treatment group is comprised of the emails from that political affiliation whose emails were marked more as spam.
For example, for Gmail, we considered emails from the right candidates as the treatment group and the emails from the left candidates as the control group because more emails from right were classified by Gmail as spam compared to the emails from left.
The reason behind considering emails from the disadvantaged political affiliation (\ie, political affiliation for which a larger percentage of emails was marked as spam) as the treatment group is that our goal is to determine whether that group of emails has been \textit{treated} unfairly by the SFA compared to the emails from the other affiliation.
Table \ref{tab:treatmentCondition} summarizes which political affiliation's emails we considered as the treatment group and which affiliation's as the control group in applying PSM on each email service.
\begin{table}[h]
	\vspace{-0.01in}
	\small
	\caption{Assignment of emails to treatment group ($Z=1$) and control group ($Z=0$) for the three services.}
	\vspace{-0.1in}
	\label{tab:treatmentCondition}
	\begin{tabular}{lccc}
		\hline
		\textit{Z} &    \textbf{Gmail} & \textbf{Outlook} & \textbf{Yahoo} \\ \hline
		1 & Right & Left & Left \\ 
		0 & Left & Right & Right \\
		\hline
	\end{tabular}
\end{table}

\subsubsection{Selection of Covariates}
The covariates whose values PSM uses should be the features that the SFAs use in determining whether any given email is spam or not.
Unfortunately, none of the three email services providers in our study have publicly disclosed what these feature are.
However, researchers have studied SFAs in the past and have identified five types of features that appear to influence the decisions of SFAs \cite{spam_filtering_ml_survey,spam_survey}.
These include
1) the meta data about email content,
2) the actual content of the email,
3) the attributes of the sender,
4) the reaction of the recipient (such as reading an email, replying to it, marking an email as spam or not-spam), and 
5) the demographics of the recipient.
Among these five types of features, the values for the last two types of features are not determined by the senders rather by the receivers (\ie, us), and thus do not need PSM.
We will analyze the impacts of the reactions of the recipients in \S \ref{subsec:interaction} and of their demographics in \S \ref{subsec:ImpactofDemographics}.
The values of the first three types of features, however, are beyond our control as they depend on who is sending the emails and what content are they including in the emails.
Thus, these three types of features require the use of PSM.

For the third type of features, \ie, the attributes of the sender, the only information we have is the IP address of the SMTP server used by the sender.
Our analysis revealed that over 80\% of the emails from the right were sent using just four digital marketing organizations, namely BlueHornet \cite{bluehornet}, Acoustic \cite{acoustic}, Amazon-SES \cite{amazonses}, and MailGun \cite{mailgun}.
Similarly, over 80\% of the emails from the left were sent using Blue State Digital \cite{bluestate}, NGP VAN \cite{ngpvan}, Amazon-SES \cite{amazonses}, and MailGun \cite{mailgun}.
As these digital marketing organizations are among the largest in the world, it is highly unlikely that the SFAs would mark emails as spam just because they were sent using one of their SMTP servers.
Thus, we do not perform PSM using the IP address of the sender's SMTP server as a covariate.

This leaves us with the first two feature types.
Next, we describe the covariates that we selected for these two types of features.

\noindent\textbf{T1: Meta Data based Covariates.}
The meta data based covariates capture the properties of the contents of the email instead of the actual contents of the email.
We calculated values for ten meta data covariates, listed and defined in Table \ref{tab:covariates}, from each email.
In selecting these covariates, one of the properties that we considered was that the distribution of any given covariate in the unmatched treatment group should have some overlap with the distribution of that covariate in the unmatched control group.
A covariate whose distributions do not have any overlap in the two groups would cause the propensity scores of the otherwise similar emails across the two groups to become very different.
This could keep PSM from creating good email pairs (see \S \ref{subsubsec:PSM--A Primer} to note that each pair has one email from unmatched treatment group and the other email from the unmatched control group such that the difference between the values of their covariates is under a threshold).

\begin{table}[htbp]
	\small
	\centering
	\caption{The ten T1 Covariates}
	\label{tab:covariates}
	\begin{tabularx}{\linewidth}{p{1.5cm}p{6.6cm}}
		\hline
		\textbf{Covariate} &    \textbf{Description} \\
		\hline Content Lexicon & The number of words in the text of the email.\\
		\hline \# Sentences & The number of sentences in the text of the email. \\ 
		\hline Readability Score & Calculated using the widely used Gunning Fog index \cite{gfog, gfog_news,gfog_singer}.\\
		\hline Social Media &  The number of times social media platforms are mentioned in the text body.\\
		\hline Thread &  Whether or not the subject starts with "Re:" or "Fwd:"\\
		\hline Upper Case & The number of upper case words in the subject and body. \\
		\hline Special Characters & The number of special characters such as !, @, * in the subject and the content. \\
		\hline \# HREFs& The number of HREFs present in the raw body of the email. \\
		\hline \# HTTP & The number of HTTP(s) links in the raw email body.\\
		\hline \# Images & The number of images referred or attached in the raw body.\\
		\hline
	\end{tabularx}
\end{table}

\noindent\textbf{T2: Content based Covariates.}
To apply PSM on emails based on their contents, we need to create pairs of matched groups such that the topics of the emails in any given pair of matched groups are closely related and the text of the emails in that pair of matched groups has similar terms.
To obtain covariates that can result in such pairs of matched groups, we use results from a recent study on election emails~\cite{elec2020}.
In this study, the authors applied a structural topic model on the content of over 105K emails and determined that there are 65 unique topics that political campaign emails are about.
They partitioned these 65 topics into six categories:
C1) political campaigns and events (topics such as Trump MAGA, primary debate, \etc), 
C2) political issues (LGBTQ, guns, \etc), 
C3) voter mobilization (winning races, voting, \etc), 
C4) explicit fundraising (donations, fundraising deadlines, \etc), 
C5) ceremonial niceties (social media, holiday wishes, \etc), and
C6) miscellaneous (signing petitions, surveys, \etc).
For each topic, the authors presented 15 FREX terms (terms that are frequent as well as exclusive for a given topic) and 15 probability terms (terms with the highest likelihood of appearing in an email on a given topic).
For the complete list of topics as well as the FREX and the probability terms for each topic, we refer the interested readers to Table 2 in~\cite{elec2020}. 

In our PSM analysis where we consider the contents of the emails, we create six pairs of matched groups, one pair per category.
To create a pair of matched groups corresponding to any given category, we use the sum of the frequency of the FREX terms and the sum of the frequency of the probability terms of each topic in any given email as covariates.
To clarify with an example, consider a hypothetical category that has four topics.
For this category, we will have eight covariates, two per topic (one from FREX terms and the other from probability terms).
The value of the covariate from FREX terms for any given topic from any given email is calculated by counting the number of times the FREX terms for that topic appear in the content (subject and body) of that email.
The value of the covariate from probability terms is calculated in the same way by counting the number of appearances of the probability terms.

\subsubsection{Applying the PSM}\label{subsubsec:ApplyingthePSM}
Next, we apply PSM to create the matched treatment and control groups of emails.
We emphasize that we do not use all the covariates of the two types (\ie, T1 and T2) together.
We separately apply PSM on T1 covariates and on the covariates of each of the six categories in T2.
To make the paper self-conatined, we have presented the technical details of how
our implementation of PSM worked in the supplementary material, \S \ref{subsubsec:ApplyingPSMonEmailsDataSet}.

\subsubsection{Observations from the Matched Groups} \label{lab:ObservationsfromtheMatchedGroups}

\begin{figure}[htbp]
	\begin{center}
		{\includegraphics[width=1\columnwidth]{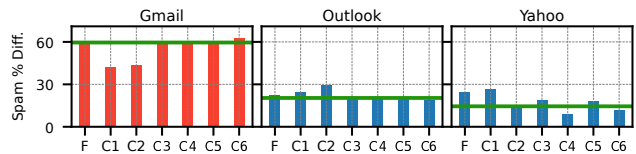}}
	\end{center}
	\vspace{-0.15in}
	\caption{\textit{Treatment Spam \%} $-$ \textit{Control Spam \%}. The color of any given bar represents the political affiliation of the emails in the corresponding treatment group.}
	\label{fig:meanDiff}
	\vspace{-0.175in}
\end{figure}

Now that we have created the matched groups, we study whether the SFAs demonstrate similar biases in the matched groups as we saw in \S \ref{subsubsec:aggregatePolBias}. Fig. \ref{fig:meanDiff}  plots the difference between the percentage of emails marked as spam in the matched treatment group and the percentage of emails marked as spam in the matched control group (\ie, \textit{Treatment Spam \%} $-$ \textit{Control Spam \%}) for each of the three email services and for each of the 7 matched groups (one matched group generated using T1 covariates and six matched groups generated using the covariates of the six categories in T2.)
Recall from Table \ref{tab:treatmentCondition} that the treatment group for Gmail is comprised of the emails from right candidates, while for Outlook and Yahoo, it is comprised of emails from the left candidates. 
For comparison, Fig. \ref{fig:meanDiff} also plots the absolute difference (green line) between the percentage of emails marked as spam in the unmatched treatment group and the percentage of emails marked as spam in the unmatched control group for each of the three email services.

From this figure, we observe that the aggregate trend that we observed in \S \ref{subsubsec:aggregatePolBias} holds even in the matched groups and the values of \textit{Treatment Spam \%} $-$ \textit{Control Spam \%} in the matched groups are fairly close to those observed in the unmatched groups (shown with green lines).
In Gmail, for the matched groups obtained using C1 and C2, we observed a 17.1\% and 16.2\% decrease, respectively.
This happened because in the unmatched groups, Gmail was marking almost all the emails on the topics of \textit{Lindsey Graham} and \textit{South Carolina} in C1 and on \textit{radical left} in C2 sent by the candidates from the right as spam.
Since there were very few emails on these topics from the left candidates, the matched groups did not contain many emails on these topics, which helped reduce the value of \textit{Treatment Spam \%}.
Nonetheless, while there was a decrease, the values of \textit{Treatment Spam \%} $-$ \textit{Control Spam \%} were still $>40\%$.

\vspace{-0.05in}
\subsection{Impact of Interactions} \label{subsec:interaction}
In this section, we study whether the interactions of the users with their email accounts cause the biases to decrease or increase.
For this, we analyze the emails collected during the Interaction Experiment E2 (\S \ref{subsubsec:e2}) for the three different types of interactions that we performed on the campaign emails:
1) reading all emails,
2) moving all emails in inbox to spam folder (\textcmt{I$\to$S}), and
3) moving all emails in spam folder to inbox (\textcmt{S$\to$I}).
We started the reading interaction with the campaign emails on August 3, 2020 and repeated daily, and the \textcmt{I$\to$S} and \textcmt{S$\to$I} interactions on September 13, 2020 and repeated every 5 days.
Recall from \S \ref{subsubsec:e2} that we performed these three interactions on three different sets of email accounts.

To study the impact of these interactions on the political biases in the SFAs, we present the observations from this interaction experiment (E2) relative to the observations from the baseline experiment (E1).
We measure this impact in two ways.
First, to observe any changes in the percentages of emails marked as spam as a result of the interactions, we compute the difference between the left (right) spam percentage in E2 and the left (right) spam percentage in E1.
Second, to analyze whether the interaction increased or decreased the political bias in the SFAs, we compute the Bias-Index ($\bi$) defined as:
\vspace{-0.1in}
\begin{equation*}
	\textbf{\textit{BI}} = \frac{|\textit{E2 Left Spam }\% - \textit{E2 Right Spam }\%|}{|\textit{E1 Left Spam }\% - \textit{E1 Right Spam }\%|}
\end{equation*}
The values of $\bi$ can lie in three ranges, interpreted as:
\begin{itemize}
	\item 0 $<$ $\bi$ $<$ 1: the bias in E2 dropped lower than the bias in E1 in response to the given interaction.
	\item $\bi$ = 1: the bias in E2 stayed the same as E1.
	\item $\bi$ $>$ 1: the bias in E2 increased compared to E1.
\end{itemize}

\subsubsection{Reading All Emails (Inbox + Spam)}
\begin{figure}[htbp]
	\vspace{-0.15in}
	\begin{center}
		{\includegraphics[width=1\columnwidth]{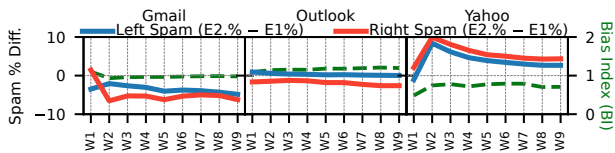}}
	\end{center}
	\vspace{-0.15in}
	\caption{Impact of reading interaction on the left and right spam percentages (left y-axis) and the political bias index (right y-axis).}
	\label{fig:read_diff}
	\vspace{-0.175in}
\end{figure}

When a user reads political emails, the spam percentage should decrease because, arguably, the user is showing interest in the received content.
However, we observed that the SFAs in the three email services reacted differently to the reading interaction.
Fig. \ref{fig:read_diff} presents the impact, at weekly intervals, of reading all emails.
The negative values of spam percentage difference in this figure demonstrate a decrease in spam percentage relative to the baseline experiment (E1) and vice versa.
We observe from this figure that in Gmail, the spam percentage marginally decreased for both the left and the right emails while still maintaining Gmail's leaning towards the left.
Due to only a marginal impact, the $\bi$ stayed approximately at 1.
In Outlook, the percentage of right spam kept decreasing over time while that of left spam stayed unchanged.
This increased the right-leaning of Outlook further, which resulted in a slightly increasing trend in $\bi$ overtime.
In Yahoo, we observed a counter-intuitive trend: the spam percentage of both the left and the right emails slightly increased due to the reading interaction.
The increase was more for right emails compared to the left emails, which resulted in slight decrease in the $\bi$.
Nonetheless, the changes in the spam percentages across all three services were minimal.

\begin{figure}[htbp]
	\vspace{-0.15in}
	\begin{center}
		{\includegraphics[width=1\columnwidth, height=0.8in]{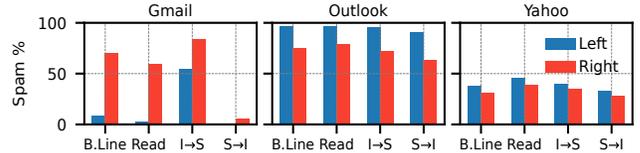}}
	\end{center}
	\vspace{-0.15in}
	\caption{Percentage of left and right emails marked as spam in baseline experiment and after the reading, \textcmt{I$\to$S}, and \textcmt{S$\to$I} interactions.}
	\label{fig:i2sFinalBiasness}
	\vspace{-0.175in}
\end{figure}

To see the net impact of the interactions on the bias, for each email service, Fig. \ref{fig:i2sFinalBiasness} shows the average percentage of the left emails and of the right emails marked as spam in the baseline experiment, after all reading interactions, after all \textcmt{I$\to$S} interactions, and after all \textcmt{S$\to$I} interactions.
This figure makes it clear that after the reading interactions, for all three email services, while there are minor changes in the percentages of the left and the right emails marked as spam, when compared to the baseline experiment, the magnitudes of the changes are negligible.
Thus, the reading interaction did not have any significant impact on the political bias of any of the three email services.

\subsubsection{Moving All Inbox Emails to Spam Folder}
\begin{figure}[htbp]
	\vspace{-0.15in}
	\begin{center}
		{\includegraphics[width=1\columnwidth]{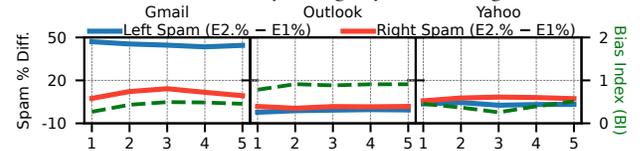}}
	\end{center}
	\vspace{-0.15in}
	\caption{Impact of \textcmt{I$\to$S} interaction on the left and right spam percentages and the bias index. X-axis shows \textcmt{I$\to$S} interaction number.}
	\label{fig:m2s_diff}
	\vspace{-0.175in}
\end{figure}

Fig. \ref{fig:m2s_diff} presents the impact of the \textcmt{I$\to$S} interaction on spam percentage and political bias index for the three services.
We observe from this figure that Gmail starts marking a significantly higher percentage of left emails as spam in response to the \textcmt{I$\to$S} interaction, reducing its left-leaning.
The percentage of right emails marked as spam also increased.
These increases are intuitive because when a user moves certain emails to the spam folder, the user is expressing that, in future, such emails should not appear in the inbox.
The increase in left spam percentage was significantly higher (by 45\%) compared to the right spam percentage (by 11\%), which is also intuitive because, in the baseline experiment, the percentage of right spam emails in Gmail was already a lot higher than the percentage of left spam emails.
Consequently, the $\bi$ of Gmail reduced significantly, showing that the \textcmt{I$\to$S} interaction significantly increased the fairness of Gmail towards the left and right emails when compared to the observations from the baseline experiment (E1).
Fig. \ref{fig:i2sFinalBiasness} shows that after all five \textcmt{I$\to$S} interactions, Gmail marked 54.2\% and 83.9\% emails from left and right, respectively, as spam across the 4 email accounts assigned to the \textcmt{I$\to$S} interaction for Gmail in E2.
While Gmail's biasness reduced, it still stayed slightly left-leaning.

Yahoo demonstrated a similar trend where the spam percentage of both left and right emails increased.
As the increase in the right spam percentage was more than the left spam percentage compared to the baseline experiment, Yahoo's $\bi$ improved (\ie, decreased) in response to the \textcmt{I$\to$S} interaction.
Fig. \ref{fig:i2sFinalBiasness} shows that after the \textcmt{I$\to$S} interactions Yahoo became almost unbiased.
Outlook, however, did not show a significant impact on the percentage of emails marked as spam.
Consequently, Outlook's $\bi$ improved only marginally in response to the \textcmt{I$\to$S} interaction.
This was expected as Outlook was already marking most left (96\%) as well as right (81\%) emails as spam (Fig. \ref{fig:dist_e1}), and thus the room for marking more emails as spam was relatively small.

To summarize, due to the \textcmt{I$\to$S} interaction, the political bias in all services improved, significantly in Gmail, moderately in Yahoo, while only marginally in Outlook.

\subsubsection{Moving All Spam Emails to Inbox}
\begin{figure}[htbp]
	\vspace{-0.15in}
	\begin{center}
		{\includegraphics[width=1\columnwidth, height=0.8in]{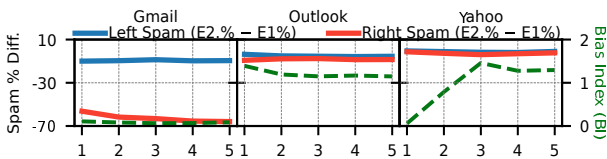}}
	\end{center}
	\vspace{-0.15in}
	\caption{Impact of \textcmt{S$\to$I} interaction on spam percentages and $\bi$. X-axis shows \textcmt{S$\to$I} interaction number.}
	\label{fig:m2i_diff}
	\vspace{-0.175in}
\end{figure}

When a user moves emails from spam to inbox, the spam percentage should decrease because the user is showing interest that such emails should appear in the inbox.
The response of Gmail to the \textcmt{S$\to$I} interaction follows this intuition while that of Outlook and Yahoo does not.
Fig. \ref{fig:m2i_diff} shows how much the spam percentage and bias index for the three services changed after each \textcmt{S$\to$I} interaction.
The negative values for both the left and right emails for Gmail show that Gmail starts putting a higher percentage of emails from both sides in inbox after just the first \textcmt{S$\to$I} interaction.
Fig. \ref{fig:i2sFinalBiasness} shows that after the five \textcmt{S$\to$I} interactions, on average, Gmail marks just 5.34\% of the right emails as spam (compared to 67.6\% in the baseline experiment) and 0\% of the left emails as spam (compared to the 8.2\% in the baseline experiment).
Thus, Gmail still maintains its left leaning, but not very strongly anymore ($\bi$ also dropped significantly).

The response of Outlook to the \textcmt{S$\to$I} interaction was counter-intuitive.
Although Outlook marked the largest number of both left and right emails as spam in the baseline experiment, its reduction in spam percentages in response to the \textcmt{S$\to$I} interaction was only marginal for both the left (by 5\%) and the right (by 8.3\%) emails.
This resulted in an increase in Outlook's right-leaning further ($\bi$ increased).
Yahoo demonstrated similar behavior as Outlook: while it marginally decreased its spam percentage, the decrease was slightly more for the right emails compared to the left emails, increasing its right-leaning ($\bi$ increased here as well).

To summarize, due to the \textcmt{S$\to$I} interaction, the political bias in Gmail reduced significantly.
However, unexpectedly, it increased in both Outlook and Yahoo because neither of the two services reacted noticeably to user's desire to not mark the emails as spam that the two services were marking as spam.

\subsection{Impact of Demographics}
\label{subsec:ImpactofDemographics}
Recall from \S \ref{subsec:EmailsAndDemographics} that our email accounts are comprised of multiple combinations of three age groups, five ethnicities, and two genders.
We observed from our data that neither the age group, nor the ethnicity, nor the gender of the account holder had any impact on how SFAs treated the emails.
Due to this and due to the space constraints, we have not shown any corresponding figures.

\section{Concluding Discussion} \label{sec:ConcludingDiscussion}

In this paper, we conducted a large-scale measurement study to examine biases in SFAs of Gmail, Outlook, and Yahoo during the 2020 US  elections.
We subscribed to a large number of left and right Presidential, Senate, and House candidates using several email accounts on the three email services.
Next, we first summarize the answers that our analysis has revealed to the four questions we posed in \S \ref{lab:introduction}, and after that provide a concluding discussion to wrap up the study.

\noindent\textbf{Summary.}
Our observations in \S \ref{subsec:polticalBiasE1} to answer \textbf{Q1} revealed that all SFAs exhibited political biases in the months leading upto the 2020 US elections.
Gmail leaned towards the left (Democrats) whereas Outlook and Yahoo leaned towards the right (Republicans).
Gmail marked 59.3\% more emails from the right candidates as spam compared to the left candidates, whereas Outlook and Yahoo marked 20.4\% and 14.2\% more emails from left candidates as spam compared to the right candidates, respectively. 

Our analysis in \S \ref{subsec:ImpactofPoliticalAffiliation} to answer \textbf{Q2} showed that the aggregate biases that we observed in the complete email data set persisted even when we considered only those emails from the left and right candidates that had very similar covariates.
Thus, it appears that the political affiliation of the sender plays role in getting an email marked as spam.

Our observations in \S \ref{subsec:interaction} to answer \textbf{Q3} showed that Gmail responded significantly more rapidly to user interactions compared to Outlook and Yahoo.
While the political bias in Gmail stayed unchanged after the reading interaction, it decreased significantly due to the \textcmt{I$\to$S} and \textcmt{S$\to$I} interactions.
Contrary to this, the political bias in Outlook, increased due to the reading and \textcmt{S$\to$I} interactions while it remained almost unchanged after the \textcmt{I$\to$S} interaction.
In Yahoo, the bias decreased due to the reading and \textcmt{I$\to$S} interactions while it increased due to the \textcmt{S$\to$I} interaction.
While the political biases changed in response to various interactions, Gmail maintained its left leaning while Outlook and Yahoo maintained their right leaning in all scenarios.

Finally, our observations in \S \ref{subsec:ImpactofDemographics} to answer \textbf{Q4} showed that the demographic factors, including age, ethnicity, and gender, did not influence the political bias of SFAs.

\noindent\textbf{Discussion.}
We conclude the paper with three thoughts.
The observation that the aggregate trend that we observed in \S \ref{subsubsec:aggregatePolBias} for the unmatched groups showed up in the matched groups as well is rather worrying because this implies that the SFAs of email services do have quite a noticeable bias.
They mark emails with similar features from the candidates of one political affiliation as spam while do not mark similar emails from the candidates of the other political affiliation as spam.
Thus the political affiliation of the sender appears to play some role towards the decision of the SFAs.
Arguably, there is also this possibility that the SFAs of email services learnt from the choices of some voters marking certain campaign emails as spam and started marking those/similar campaign emails as spam for other voters.
While we have no reason to believe that there were deliberate attempts from these email services to create these biases to influence the voters, the fact remains there that their SFAs have learnt to mark more emails from one political affiliation as spam compared to the other.
As these prominent email services are actively used by a sizable chunk of voting population and as many of the voters today rely on the information they see (or don't see) online, such biases may have an unignorable impact on the outcomes of an election.
It is imperative for the email services to audit their SFAs to ensure that any properties of the sender that they consider in determining whether any given email is spam or not are not, unknowingly, putting one side at an advantage compared to the other.

Second, the general perception is that when a user reads emails, marks them as spam, or moves them from the spam folder to inbox, the SFA adapts to user's preferences.
While our observations agree with this perception, very strongly for Gmail and to a smaller extent for Outlook and Yahoo, this adaptation does not necessarily eliminate the political bias.
Some interactions do reduce the bias, but that effect is not consistent across all the email services.
In other words, we did not find any consistent actions that one could recommend to users to help them reduce the bias in the way the SFA treats political emails that are sent to them.

Third, if an undecided voter receives too many emails from one political party, there is a likelihood that they may get swayed towards that party.
As users open their spam folders very infrequently, it is unlikely that most undecided voters will undertake the effort to open the spam folder and mark some campaign emails as not-spam to make the SFA unbiased.
Therefore, it is important for the SFAs to be unbiased at the outset without relying on explicit user feedback.

We conclude by noting that fairness of spam filtering algorithm is an important problem that needs dedicated attention from email service providers, particularly due to the enormous influence that electronic communication has in our lives today and will have going forward.
However, it is not an easy problem to solve.
Attempts to reduce the biases of SFAs may inadvertently affect their efficacy. 
Therefore, there is an imminent need to develop techniques that reduce the biases of SFAs while simultaneously ensuring that the users do not receive unwanted emails.

\bibliographystyle{ACM-Reference-Format}
\bibliography{bibliography}

\appendix
\section{Supplementary Material}
\subsection{PSM -- A Primer}\label{subsubsec:PSM--A Primer}
To create the matched treatment and control groups, PSM performs two steps.
In the first step, it calculates a propensity score for each member of the unmatched treatment and control groups using the values of the covariates of that member.
Propensity score for any given member is defined as the probability that that member belongs to the treatment group conditioned on the member's covarites.
Formally, for any member $i$, let $Z_i$ represent the indicator random variable that is 1 if the member $i$ belongs to the treatment group and 0 if the member $i$ belongs to the control group.
Let the number of covariates that can be observed for any given member be $K$ and let $X_{ij}$ represent the $j$\textsuperscript{th} covariate of the $i$\textsuperscript{th} member.
The propensity score for the $i$\textsuperscript{th} member, represented by $e_i$, is then given by the following equation.
\begin{equation*}
	e_i = Pr(Z\textsubscript{i} = 1 | X_{i1}, X_{i2}, \ldots, X_{iK})
	\label{equ:propScore}
\end{equation*}

In the second step, PSM uses the scores that it calculated for all the members of the unmatched treatment and control groups, and creates matched treatment and control groups.
More specifically, for each member $i$ of the unmatched treatment group, PSM finds a member $j$ in the unmatched control group such that the absolute difference between the scores of the two members (\ie, $|e_i-e_j|$) is below a certain threshold.
If it is able to find such a member $j$ in the control group, it puts the member $i$ of the unmatched treatment group into the matched treatment group and the member $j$ of the unmatched control group into the matched control group.
If it is not able to find a member in the control group for which $|e_i-e_j|$ is below the threshold, it discards the member $i$ of the unmatched treatment group and moves on to member $i+1$ and repeats the step above.
After iterating through all the members of the unmatched treatment group, PSM returns the new treatment and control groups, where the distribution of any given covariate is very similar in the two matched groups.

\subsection{Applying PSM on Emails Data Set}\label{subsubsec:ApplyingPSMonEmailsDataSet}
Next, we describe how we apply the two steps of PSM (mentioned in \S \ref{subsubsec:PSM--A Primer}) to create the matched treatment and matched control groups of emails from the unmatched treatment and unmatched control groups.
We emphasize that we do not use all the covariates of the two types described above together.
Instead, we separately apply the two steps of PSM on T1 covariates and on the covariates of each of the six categories in T2.

\noindent\textbf{Step 1: Propensity Score Estimation.}
To estimate the propensity score for each email, we first create a logistic regression model using covairates as the independent variables and $Z$ as the dependent variable.
Recall from \S \ref{subsubsec:PSM--A Primer} and Table \ref{tab:treatmentCondition} that $Z$ is equal to 1 for any email that belongs to the treatment group and is equal to 0 for emails that belong to the control group.
As different covariates can have different amounts of correlations with $Z$, we incorporate lasso regularization when creating the logistic regression \cite{categorical,lasso}.
Lasso regularization discounts the effect of covariates that do not have a noticeable correlation with $Z$.
After creating the logistic regression model, for each email, we feed its covariates to this model and the model outputs a value between 0 and 1, which is the propensity score for the email.
The method that we just described is one of the most common methods in literature to estimate the propensity scores \cite{psmbook,austin2011}.

\begin{figure}[t]
	\begin{center}
		{\includegraphics[width=1\columnwidth]{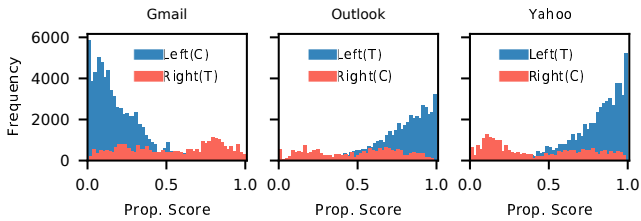}}
	\end{center}
	\caption{Histograms of the propensity scores of the emails in the unmatched treatment (T) and the unmatched control (C) groups using T1 covariates.}
	\label{fig:pscore_hist}
	\vspace{-0.075in}
\end{figure}

\noindent\textbf{Step 2: Matching.}
Fig. \ref{fig:pscore_hist} shows the distributions of the propensity scores in the unmatched treatment (marked as T) and control (marked as C) groups for the three services when using T1 covariates.
We observe from this figure that for each email service, there is a decent overlap between the distributions of the left and the right emails.
This shows that we should be able to create matched treatment and control groups with large enough number of emails in each group such that any observations that we make from them are statistically significant.
We made very similar observations for the T2 covariates.

As with the propensity score estimation, several methods have been proposed in literature to create matched groups.
We again employed one of the commonly used methods, known as caliper matching \cite{austin2011}.
Caliper matching not only creates well-matched pairs of emails but also excludes any emails from the treatment group for which a good match in the control group is not available.
Caliper matching matches any given email $i$ in the unmatched treatment group with that email $j$ in the unmatched control group for which the absolute difference between propensity scores of the two emails is minimum and at the same time less than a threshold $B$.
The threshold $B$, also known as the caliper width, is defined as $B = \kappa\times\sigma$ where $\sigma$ is the standard deviation of all the propensity scores in the unmatched treatment and control groups, and $\kappa$ is a multiplicative constant.
In \cite{austin2011}, the authors suggested to set $\kappa=0.2$.

The quality of the matched treatment and control groups is quantified in terms of standardized mean difference (SMD), given by the following equation:
\begin{equation*}
	\text{SMD} = (\overline{e}\textsubscript{T}-\overline{e}\textsubscript{C})/\sqrt{(s\SPSB{2}{T}+s\SPSB{2}{C})/2}
\end{equation*}
where $\overline{e}\textsubscript{T}$ and $s\SPSB{2}{T}$ represent the mean and variance of the propensity scores in the matched treatment group and $\overline{e}\textsubscript{T}$ and $s\SPSB{2}{T}$ represent the same for the matched control group.
The quality of the matched groups is considered good when the absolute value of the SMD is $\leq$ 0.1 \cite{austin2011}.

\begin{figure}[t]
	\begin{center}
		{\includegraphics[width=1\columnwidth]{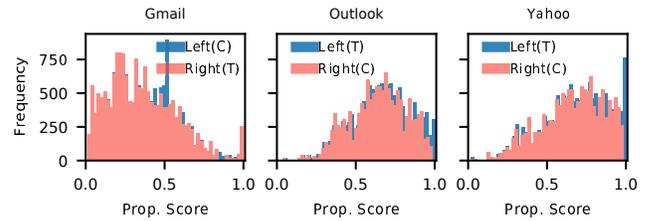}}
	\end{center}
	\caption{Histograms of the propensity scores of the emails in the matched treatment (T) and the matched control (C) groups using T1 covariates.}
	\label{fig:pscore_hist_after_matching}
	\vspace{-0.075in}
\end{figure}

To create the matched treatment and control groups, we start with $\kappa=0.2$ and find an email in the unmatched control group corresponding to each email in the unmatched treatment group as per the caliper matching criteria mentioned above.
Next, we check the quality of the resulting matched treatment and control groups by calculating the value of SMD.
If the absolute value of SMD turns out to be greater than 0.1, we tighten the caliper width by decrementing $\kappa$ in the steps of 0.02, and repeat the steps again until $|$SMD$|$ drops below 0.1.
At this point, we obtain the final matched treatment and control groups that have the same number of emails, and the distribution of the propensity scores as well as that of any covariate is very similar across the two groups.
For example, Fig. \ref{fig:pscore_hist_after_matching} shows the distribution of the propensity scores in the matched treatment and in the matched control groups for the three services using the T1 covariates.
We indeed observe that for each email service, the distribution of the propensity scores in the matched treatment group very closely overlaps with the distribution of the propensity scores in the matched control group.
This shows that the two matched groups for each email service that PSM has created for us are indeed comprised of two very similar sets of emails.
We made very similar observations for the T2 covariates.

\end{document}